## S33
## Cyber Security in Energy Informatics: A Non-technical Perspective

Duong Dang[1,*], Tero Vartiainen[1], and Mike Mekkanen[1]
University of Vaasa School of Technology and Innovations, University of Vaasa, Wolffintie 34, 65200, Vaasa, Finland

**Correspondence**: Duong Dang (duong.dang@uwasa.fi)

**Summary**
Literature in cyber security including cyber security in energy informatics are overly technocentric focuses that may miss the chances of understanding a bigger picture of cyber security measures. This research thus aims to conduct a literature review focusing on non-technical issues in cyber security in the energy informatics field. The findings show that there are seven non-technical issues have been discussed in literature, including education, awareness, policy, standards, human, and risks, challenges, and solutions. These findings can be valuable for not only researchers, but also managers, policy makers, and educators.

### Introduction
Energy transition is becoming an important trend nowadays. This trend brings cyber threats from information and communications technology (ICT) to the energy sector as ICT is embedded into energy systems. Although scholars have focused on cyber security to improve the cyber security of the energy system, the majority of literature focuses on technical aspects of cyber security. However, literature shows that technology is not enough in preventing cyber threats [1] as technical systems, the humans who operate them, and organizational contexts are all important [2].
Scholars thus call for a more holistic approach for effectiveness of security measures [1]. As a result, we conducted this study by focusing on non-technical issues in cyber security in energy informatics research. We particularly focus on the following research question: What are the main themes related to non-technical issues of cyber security in energy informatics? In this paper, aspects like awareness, policies, and organizational structures can be seen as non-technical aspects, while the specific technologies (e.g., firewalls, cyber-physical systems, algorithms) can be categorized as the technical aspects [3].

### Methods
To answer the research question, we conducted a systematic literature review in the field of cyber security in energy informatics. The search was carried out in March, 2022 by using the Scopus database. It is noted that IEEE Xplore and ACM Digital Library, which are largely Scopus-indexed. We focused on journals, conferences and book chapters with the following search string is used: (("smart grid" OR "energy informatics") AND ("cyber security" OR "cybersecurity" OR "cyber-security")) AND (LIMIT-TO ("journals" or "conferences" or "chapters") AND (LIMIT-TO (LANGUAGE, "English")). The result was 1151 papers with this search string. We then performed checking duplicated or not-relevant papers. Next, we eliminated those survey and literature review papers, as well as omitted pure technical papers. After that, we read the remaining papers in full to eliminate those papers that do not discuss issues related to the review's scope. After this step, 211 papers were selected for review. Moreover, given that Energy Informatics journal is one of the main outlets in the field, we conducted a search on this journal. Two additional papers were found. It means that 213 papers were selected for this study.

### Findings
This section presents non-technical issues that have been discussed in selected papers.
**Education**. Literature discussed two main program types of educations on cyber security in energy informatics, that are training program for professional [4] (e.g., Skill gaps, workforce, team-taught, living lab), and training program for students [5] (e.g., Curriculum, course for STEM students, courses for undergraduate students). The curriculum for cyber security is discussed in three different levels that are equally important: Cyber security for all, cyber operations, and cyber-informed engineering curriculum [6]. Moreover, pedagogical pillars for energy informatics cyber security education are active learning, project-based learning, Piaget's learn-by-doing posture, and constructivism [7].
**Awareness**. Cyber security awareness is crucial for the effectiveness of organizational security [8]. Two types of awareness are often discussed in those selected papers, including social awareness and situation awareness. Social awareness helps enhance and improve cyber security in energy informatics. The awareness includes socio-economic consumer data; load disaggregation capability; end-use device database; and smart power hub [9]. Situation awareness indicates that architecture and tools are needed to help operators monitor and be aware of actual threats that exist between the network level and the business level [10].
**Policy**. There are two categories that have been discussed in those selected papers regarding policy (including legal, and regulatory), that are policy challenges and policy itself [11]. Policy challenges regarding cyber security in energy informatics is one of the main concerns of scholars. Examples of those concerns include privacy, personal data, data security of sharing information in applications used by the smarter grid devices. Scholars also discuss unclear guidance on mandates and roles of organizations/countries on the topics of cyber security [12].
**Geography**. Majority selected articles come from the EU [13], North America [14], and Asia [15]. The NIST and the ISO frameworks/standards are widely adopted by the EU Critical Infrastructure (e.g., ISO 27001 for IT security, or ISA 62443 for Operational Technology security). The European Union Agency for Cyber Security (ENISA) and the Department of Homeland Security (DHS) in the USA provides security guidelines to support the implementation of high security standards for critical infrastructures. While the US and the EU's scholars focus on solutions, it seems that Asian researchers are interested in threats, and needs for cyber security solutions (e.g., a need for a domain specific regulatory framework in India [15]).
**Standards**. Standards (including frameworks and models) have been categorized into two types: Standards in general (e.g., Cyber security assessment [16]) and specific standards for a system (e.g., SCADA system cyber security standards [17]). Scholars discussed security assessment techniques [16], such as reviews passive, vulnerability identification, and vulnerability analysis (e.g., IEC 62351, IEC 62443, IEC 62056-5-3, ISO 15118, ISO/IEC 27019). Literature also discussed particular standards for a specific system, such as SCADA system [17].
**Humans**. Literature indicates that human behavior is considered one of the important factors in cyber security in energy informatics [18]. Two main issues of human factors have been discussed are roles of human failures in security, and cyber security leadership. For example, the security of smart technologies in the energy systems cannot rely only on technical solutions; humans play a significant role in failure in cyber security, whether it is intentional or unintentional [19]. Human aspects also include management (c.f., [20]).
**Solutions**. Solutions for cyber security in energy informatic include issues related to risks [21], challenges [22], and solutions [23]. Risk topics include risk assessment of power information control systems [24], energy internet, digital secondary substations, renewable energy, smart grids, physical systems [25], and economic risk [26]. Challenges in smart grid include, for example, hyperphysical challenges in smart grid, information security and privacy challenges, technology challenges (AI, big data, IoT) [27], and communications [28]. Solutions include solution in general, solution in cyber security in smart grid, and solution in cyber security in physical system [29].and software. Energy systems thus do no longer rely on physical and local

## Conclusion and Outlook

We have retrieved seven non-technical issues on cyber security in energy informatics, they include education, awareness, policy, geography, standard, human, and solution. From these findings, it can be argued that security awareness is one of the important issues in cyber-defense. Energy system devices are gradually replaced by standard IT protocols and commercial-of-the-shelf hardware and software. Energy systems thus do no longer rely on physical and local measures for their operations. We thus suggest that more study in assessments of maturity cyber security awareness in organizations are needed, such as awareness models or frameworks in cyber security in energy informatics.

Future research will focus on synthesizing these findings to benefit not only researchers, but also managers, policy makers, and educators. For example, identifying socio-technical gaps in the context of digital transformation [30] could help measures on preventing cybercrimes in a holistic way in cyber security in energy informatics.

## Funding
This work is funded by the University of Vaasa, Finland.

## Availability of data and materials
Not Applicable.

## Author's contributions
D.D. designed, conducted, analyzed, and drafted the manuscript. T.V. discussed and commented on the manuscript. M.M. commented on the final draft of the manuscript. All authors have read and approved the manuscript.

## References
1. Bunker G (2012) Technology is not enough: Taking a holistic view for information assurance. Inf Secur Tech Rep 17:19–25
2. Malatji M, Von Solms S, Marnewick A (2019) Socio-technical systems cybersecurity framework. Inf Comput Secur 27:233–272
3. Whitman M, Herbert M (2022) Principles of Information Security, 7th Edition. Cengage Learning, Inc.
4. Siemers B, Attarha S, Kamsamrong J, et al (2021) Modern trends and skill gaps of cyber security in smart grid: Invited paper. In: M. A (ed) 19th IEEE Int. Conf. Smart Technol. EUROCON 2021, pp 565–570
5. Kuzlu M, Popescu O, Jovanovic VM (2021) Development of a Smart Grid Course in an Electrical Engineering Technology Program. 2021 ASEE Virtual Annu. Conf. ASEE 2021
6. Loo SM, Babinkostova L (2020) Cyber-physical systems security introductory course for STEM Students. 2020 ASEE Virtual Annu. Conf. ASEE 2020 2020-June:
7. Yardley T, Uludag S, Nahrstedt K, Sauer P (2015) Developing a Smart Grid cybersecurity education platform and a preliminary assessment of its first application. 44th Annual Front Educ Conf FIE 2014.
8. Hagen JM, Albrechtsen E, Hovden J (2008) Implementation and effectiveness of organizational information security measures. Inf Manag Comput Secur 16:377–397
9. Singh A, Pooransingh A, Ramlal CJ, Rocke S (2017) Toward Social Awareness in the Smart Grid. In: G.S. T (ed) 8th Int. Conf. Comput. Intell. Commun. Networks, CICN 2016, Trinidad and Tobago, pp 537–543
10. Angelini M, Santucci G (2015) Visual cyber situational awareness for critical infrastructures. In: P. B, T. I, S. T (eds) 8th Int. Symp. Vis. Inf. Commun. Interact. VINCI 2015. ACM, Rome, Italy, pp 83–92
11. Mah D, Leung KP-Y, Hills P (2014) Smart grids: The regulatory challenges. Green Energy Technol 0:115–140
12. Mylrea M (2017) Smart energy-internet-of-things opportunities require smart treatment of legal, privacy and cybersecurity challenges. J World Energy Law Bus 10:147–158
13. Genzel C-H, Hoffmann O, Sethmann R (2017) IT-security for smart grids in Germany: Threats, countermeasures and perspectives. In: M. S, N.-A. L-K (eds) 16th Eur. Conf. Cyber Warf. Secur. ECCWS 2017. Curran Associates Inc., University of Applied Sciences Bremen, Germany, pp 137–145
14. Baggott SS, Santos JR (2020) A Risk Analysis Framework for Cyber Security and Critical Infrastructure Protection of the U.S. Electric Power Grid. Risk Anal 40:1744–1761
15. Ananda Kumar V, Pandey KK, Punia DK (2014) Cyber security threats in the power sector: Need for a domain specific regulatory framework in India. Energy Policy 65:126–133
16. Leszczyna R (2018) Standards on cyber security assessment of smart grid. Int J Crit Infrastruct Prot 22:70–89
17. Sommestad T, Ericsson GN, Nordlander J (2010) SCADA system cyber security - A comparison of standards. IEEE PES Gen Meet PES 2010.
18. Back S, LaPrade J (2019) The Future of Cybercrime Prevention Strategies: Human Factors and A Holistic Approach to Cyber Intelligence. Int J Cybersecurity Intell Cybercrime 2:1–4
19. Aldabbas M, Teufel B (2016) Human aspects of smart technologies' security: The role of human failure. J Electron Sci Technol 14:311–318
20. D. Dang and T. Vartiainen, "Changing Patterns in the Process of Digital Transformation Initiative in Established Firms: The Case of an Energy Sector Company," PACIS 2020 Proc., Jun. 2020.
21. Christensen D, Martin M, Gantumur E, Mendrick B (2019) Risk Assessment at the Edge: Applying NERC CIP to Aggregated Grid-Edge Resources. Electr J 32:50–57
22. de Kinderen S, Kaczmarek-Heß M (2021) Making a Case for Multi-level Reference Modeling – A Comparison of Conventional and Multi-level Language Architectures for Reference Modeling Challenges. In: F. A, R. S, S. S (eds) Int. Conf. Wirtschaftsinformatik, WI 2021. Springer. pp 342–358
23. B. Eltahawy and D. Dang, "Understanding Cyberprivacy: Context, Concept, and Issues," Wirtschaftsinformatik 2022 Proc., Jan. 2022.
24. M. Mekkanen, T. Vartiainen, and D. Dang, "Develop a Cyber Physical Security Platform for Supporting Security Countermeasure for Digital Energy System," Scand. Simul. Soc., vol. 185, pp. 219–225, Mar. 2022.
25. Al Zadjali A, Ali S, Balushi TA (2016) Risk assessment for cyber-physical systems: An approach for smart grid. In: K.S. S (ed) 27th Int. Bus. Inf. Manag. Assoc. Conf. - Innov. Manag. Educ. Excell. Vis. 2020, pp 3204–3213
26. Hébert C (2013) The most critical of economic needs (risks): A quick look at cybersecurity and the electric grid. Electr J 26:15–19
27. Younes Z, Alhamrouni I, Mekhilef S, Khan MRB (2021) Blockchain Applications and Challenges in Smart grid. In: 5th IEEE Conf. Energy Conversion, CENCON 2021, Kuala Lumpur, Malaysia, pp 208–213
28. Srivastava I, Bhat S, Singh AR (2022) Smart Grid Communication: Recent Trends and Challenges. Lect Notes Electr Eng 824:49–75
29. Velayutham Y, Bakar NAA, Hassan NH, Samy GN (2021) IoT security for smart grid environment: Issues and solutions. Jordanian J Comput Inf Technol 7:13–2430.
30. D. Dang and T. Vartiainen, "Digital strategy patterns in information systems research," PACIS 2019 Proc., Jun. 2019.